\documentclass[aps,prd,twocolumn,nofootinbib,showpacs,superscriptaddress]{revtex4-1}

\usepackage{amssymb}
\usepackage{graphicx}
\usepackage{amsmath, amsthm}
\usepackage{epstopdf}
\usepackage{hyperref}

\newcommand{\be}{\begin{equation}}
\newcommand{\ee}{\end{equation}}

\begin{document}

\title{Covariantizing the interaction between dark energy  and dark
matter} 

%\author{} 
%\email[]{vfaraoni@ubishops.ca}
%\affiliation{Physics Department and STAR Research Cluster, 
%Bishop's University\\
%Sherbrooke, Qu\'ebec, Canada J1M~1Z7}

\author{Valerio Faraoni}
\email{vfaraoni@ubishops.ca}
\affiliation{Physics Department and STAR Research Cluster, 
Bishop's University,\\
Sherbrooke, Qu\'ebec, Canada J1M~1Z7
}
\author{James B. Dent}
\email{jbdent@louisiana.edu}
\affiliation{Department of Physics, University of Louisiana at Lafayette,
Lafayette, LA 70504-4210, USA}

 \author{Emmanuel N. Saridakis}
\email{Emmanuel\_Saridakis@baylor.edu}
 \affiliation{Physics Division, National Technical University of Athens,
15780 Zografou Campus,  Athens, Greece}
\affiliation{Instituto de F\'{\i}sica, Pontificia Universidad Cat\'olica
de Valpara\'{\i}so, Casilla 4950, Valpara\'{\i}so, Chile}

%\date{\today}
\begin{abstract} 
Coupling dark energy and dark matter through an effective fluid description
is a very common procedure in cosmology, however it always remains in
comoving coordinates in the special FLRW space. We construct a 
consistent,
general, and covariant formulation, where the interaction is a natural
implication of the imperfectness of the fluids. This imperfectness makes
difficult the final step towards a robust formulation of interacting fluids,
namely the construction of a Lagrangian whose variation would give rise to
the interacting equations. Nevertheless, we present a formal solution to this
problem for a single fluid, through the introduction of an effective
metric. 
\end{abstract}

% insert suggested PACS numbers in braces on next line
\pacs{ 98.80.-k, 95.36.+x, 95.35.+d, 47.10.-g }
% insert suggested keywords - APS authors don't need to do this
\keywords{}

\maketitle
% If in two-column mode, this environment will change to single-column
% format so that long equations can be displayed. Use
% sparingly.
%\begin{widetext}
% put long equation here
%\end{widetext}

\section{Introduction}
\label{sec:1}

Dark energy and dark matter are the basic constituents of the universe
\cite{Ade:2013zuv}. Although there are theories postulating that 
they may correspond to a
unified ``dark sector''(for instance in Chaplygin-gas-like theories
\cite{Kamenshchik:2001cp}), detailed cosmological observations, and
especially the clustering properties of dark matter  
\cite{Scoville:2006vr,Kilbinger:2012qz,Anderson:2013zyy} 
in contrast with the homogeneity of dark energy, suggest with a 
great certainty that dark
energy and dark matter are two separate sectors.
%(the expected
%``discovery'' of dark matter in the relatively near future will be the
%final verification on this). 
Hence, one could construct scenarios in which the dark energy and 
dark matter
sectors interact \cite{Billyard:2000bh}, since this interaction, apart from
being theoretically allowed, could have an important phenomenological
implication, namely alleviating the coincidence problem ({\em 
i.e.}, why are the
current dark energy and matter densities of the same order although they
evolve differently).

In the existing literature the interaction is described in a very 
simple way, that is with the arbitrary modification of the 
equations of motion. In particular, one handles both dark energy 
and dark matter as perfect fluids in the framework of General 
Relativity,\footnote{We mention that in the framework of modified 
gravity one can obtain interactions of dark matter with the extra 
degrees of freedom of gravitational modification which play the 
role of an effective dark energy, through the transformation to the 
Einstein frame, but this is a completely different issue 
\cite{Sotiriou:2008rp,DasAlMamon}.} whose total 
conservation is arbitrarily 
split into non-conserved ``interacting'' parts:
\begin{eqnarray}
\label{conservtotal}
&&\nabla^b T_{ab}^{(tot)}=
\nabla^b\left(T_{ab}^{(DM)}+T_{ab}^{(DE)}\right)=0\\
&&\ \ \ \ \ \ \ \ \ \  \ \   \ \ \ \  \ \  \Rightarrow\nonumber\\
&&\nabla^b T_{ab}^{(DM)}=Q_a \ \, \text{and}\,\ \nabla^b
T_{ab}^{(DE)}=-Q_a \,,
\label{conservsepar}
\end{eqnarray}
where the quantity $Q_a$ is introduced as a phenomenological descriptor of the 
interaction,  the form of which is assumed arbitrarily, too.
Although this arbitrary splitting is mathematically correct, there is not a
procedure determining how the system described in 
eqs.~(\ref{conservtotal}) and~(\ref{conservsepar}) could 
physically arise, and especially  how to determine $Q_a$ (see, 
however, Refs.~\cite{extrarefs}). 

In principle, any fundamental theory should be characterized by a 
Lagrangian whose variation gives rise to the equations of 
motion. If the
microscopic nature of dark matter and especially of dark energy were known,
one could write down a Lagrangian with all possible interaction terms, and
then varying it one could obtain the complete and exact interacting 
equations of
motion and the corresponding interaction terms, similarly to the interactions
within the Standard Model. Since such a microscopic
description is
currently impossible, one could still hope to describe the 
dark energy-dark matter interaction in an effective way, writing an 
effective
Lagrangian whose variation could give rise to (\ref{conservsepar}).  There 
has been a recent wealth of explorations of dark matter direct and 
indirect detection, as well as collider production through the means of the 
effective field theory approach (for a small sampling of the field see for
 example \cite{Harnik:2008uu,Beltran:2010ww,Fan:2010gt,Fox:2011pm,
 Fitzpatrick:2012ix,DeSimone:2013gj,Kumar:2013iva,Busoni:2013lha}).  
 Similarly there have been forays into describing dark energy and 
 modified gravity via effective theories 
\cite{Bloomfield:2012ff,Gubitosi:2012hu}.
 Nevertheless, in the existing literature regarding the coupling of dark energy and dark matter,
neither the microscopic description nor the effective field theory approach are employed, and
the relations given in (\ref{conservsepar}) are imposed by hand.

Therefore the important question that arises naturally is the following:
Is this widespread formalism consistent? In spite of being always presented
only in comoving coordinates in Friedmann-Lema\^itre-Robertson-Walker (FLRW) 
geometry, can it be given a covariant
formulation? And ideally, can we write down (effective) Lagrangians, whose
variation would give rise to (\ref{conservsepar})? This is the field of
interest of the present work.

\section{Two fluid interaction: the standard procedure}
 
In the discussions of coupled dark energy and dark matter in cosmology,
one considers two coupled fluids in a FLRW space. Let us 
restrict, for simplicity, to a spatially flat 
FLRW geometry (which is anyway the one encountered in the literature) 
described by the line element
\be 
ds^2=-dt^2 +a^2(t) \left( dx^2+dy^2+dz^2 \right) \,, 
\ee 
where $a(t)$ is the scale factor. The two fluids commonly 
considered in the literature are assumed to have energy densities 
$\rho_{1,2}$ and pressures $P_{1,2}$ depending only on time, in 
order to respect spatial homogeneity and isotropy, and apart from the 
interaction they mimic perfect fluid behaviour. They are 
usually assumed to satisfy the equations of motion
\be\label{fluid1} 
\dot{\rho}_1+3H \left(P_1+\rho_1 \right) =Q \,, 
\ee 
\be\label{fluid2} 
\dot{\rho}_2+3H \left(P_2+\rho_2 \right) =-Q \,. 
\ee 
where $H \equiv \dot{a}/a$ is the Hubble parameter and an overdot denotes
differentiation with respect to the comoving time $t$. The quantity $Q$
quantifies the interaction and its forms are considered completely
arbitrarily, with the obvious requirement to depend only on time due to
homogeneity and isotropy. The usual choices of $Q$ encountered in the
literature make this quantity proportional to $\rho_{1,2}$ or to 
the Hubble parameter $H$, and their powers
\cite{miscQ}, and one can additionally use observations in
order to constrain their forms \cite{Wang}.

The two equations (\ref{fluid1}) and (\ref{fluid2}) are concocted so that, by
adding them together, a ``total fluid'' of energy density 
\be
\rho_{tot}=\rho_1 +\rho_2 
\ee
and pressure 
\be
P_{tot}=P_1+P_2 
\ee
satisfies the conservation equation
\be \label{totalconservation}
\dot{\rho}_{tot}+3H \left(P_{tot}+\rho_{tot} \right) =0 \,. 
\ee 
Actually, as we discussed in the Introduction, the aforementioned arbitrary
splitting exactly arises from this conservation of the total fluid. 
The ``total'' fluid has effective equation of state parameter 
\be
w_{tot}(t) \equiv \frac{P_{tot}}{\rho_{tot}} = 
\frac{P_1+P_2}{\rho_1 +\rho_2} 
=\frac{w_1\rho_1 + w_2 \rho_2}{\rho_1 +\rho_2} \,,\label{weff}
\ee
namely it is an average of the equation of state parameters of the individual
fluids $w_{i}$ weighted by their energy fractions (density parameters)
$\rho_i/\rho_{tot}$. Although the individual $w_1$ and $w_2$ may both be constant, the resulting 
$w_{tot}$ is not, except for the trivial cases $w_1=w_2$ (in which case there
is a single fluid with density $2\rho$ and pressure $2P$) or constant
$\rho_1$ and $\rho_2$. Based on this formulation, a non-insignificant amount
of literature ({\em e.g.}, \cite{AmendolaTsujikawabook, miscQ,Wang,
Bertolami,
ThomasValerio}) has appeared. 
%which uses the formalism expressed by eqs.~(\ref{fluid1})-(\ref{weff}).  

\section{Two fluid interaction: a consistent covariant picture}
\label{sec:2}

Eqs.~(\ref{fluid1}) and (\ref{fluid2}) can be obtained in a  
consistently covariant picture if the two fluids are described by 
the  stress-energy tensors
\begin{eqnarray}
T^{(1)}_{ab} &= & \left( P_1 + \rho_1 \right) u_au_b +P_1 g_{ab}
+ q_au_b +q_b u_a   \,, \label{bbb}\\
&&\nonumber\\
T^{(2)}_{ab} &= & \left( P_2 + \rho_2 \right) u_au_b +P_2 g_{ab} 
-q_au_b -q_b u_a \,,
\end{eqnarray}
where $u^a$ is the common 4-velocity of the two fluids, a timelike 
unit  vector pointing in the time direction. The two fluids are not 
tilted with  respect to each other, that is, they have the same 
4-velocity $u^a$ and they  ``see'' the same 3-space orthogonal 
to $u^a$ with 3-metric $h_{ab}=g_{ab}+u_a u_b$ (${h^a}_b$ is the 
projection operator on this 3-space). $q^c$ is a current 
energy density, a timelike vector which describes the transfer of energy
between the two fluids. Due to spatial isotropy, $q^c$ cannot have any 
spatial component and must point in the time direction,
\be
q^c=\alpha(t) u^c \,,
\ee
where $\alpha $ is a function of time which must be non-negative for $q^c$ to
be future-oriented.

We mention here that the two fluids are imperfect fluids,
but not in the usual sense \cite{Stephani}. Usually, the term 
$q_au_b +q_b u_a$ in an imperfect fluid is associated with a purely 
spatial energy current density (that is, one satisfying 
$q^cu_c=0$ \cite{Stephani}), but this is not the case here: the 
flux density of 
energy must be parallel to $u^c$ in order not to violate spatial 
isotropy. Because of this, and contrary to the standard textbook 
imperfect fluid, the traces of $T^{(i)}_{ab}$ are not the same as 
those of a perfect fluid, namely
\be
T^{(i)} = -\rho_i+3P_i \mp 2\alpha  \,.
\ee
Note that the ``total'' stress-energy tensor 
\be\label{totalTmunu}
T_{ab}^{(tot)} = T_{ab}^{(1)} + T_{ab}^{(2)} 
\ee
is covariantly conserved
\be
\nabla^b T_{ab}^{(tot)} =0 \,,
\ee
and the ``total'' energy density and pressure associated with it 
are 
$\rho_{tot}=\rho_1 +\rho_2 $ and $P_{tot}=P_1+P_2 $. On the other hand, the
covariant 
divergence of the $i$-th fluid  ($i=1,2$) stress-energy tensor 
$T_{ab}^{(i)}$ is
\begin{eqnarray}
&&\nabla^b T_{ab}^{(i)} = u_a u^b \nabla_b P_i + u_a u_b
\nabla^b 
\left( \rho_i \pm 2\alpha \right)+\nabla_a P_i  \nonumber\\ 
&&\ \ \ \ \ \ \ \ \ \ \ \ \ +
\left( P_i+\rho_i \pm 2\alpha \right) u^b \nabla_b u_a  \nonumber\\ 
&&\ \ \ \ \ \ \ \ \ \ \ \ \ +\left( P_i 
+ \rho_i  \pm 2\alpha \right) u_a \nabla^b u_b  \,,
\end{eqnarray}
where the upper sign corresponds to fluid~1 and the lower one to
fluid~2. 
Projection along the time direction $u^a$ gives
\be
u^a \nabla^b T_{ab}^{(i)} = 
\left( \dot{\rho}_i \pm 2\dot{\alpha} \right) +3H\left( P_i +\rho_i 
\pm 2\alpha \right) \,,
\ee
where $\dot{\rho}_i \equiv u^a \nabla_a \rho_i$, {\em etc.} 
By imposing that $u^a \nabla^b T_{ab}^{(i)} =0$, the two fluids 
are conserved separately, but their perfect-fluid components $ 
\left( P_i+\rho_i \right) u_au_b +P_i g_{ab}$ are not, having 
non-zero covariant divergences which satisfy
\be
u^a \nabla^b \left[ \left( P_i+\rho_i \right) u_au_b +P_i g_{ab} 
\right] = \pm 2 \left( \dot{\alpha} +\alpha \nabla_b u^b \right) 
\,.
\ee
In an FLRW background this equation becomes 
\be
\dot{\rho}_i +3H \left( P_i +\rho_i \right) =\mp 2\left( 
\dot{\alpha}+3H\alpha \right) \,.
\ee
Hence, one can clearly see that imposing the right hand side of this
equation to be equal to $\pm 
Q$, eqs.~(\ref{fluid1}) and (\ref{fluid2}) are reproduced. In this case
$\alpha$ and $Q$ satisfy the relation
\be
\dot{\alpha}+3H\alpha+\frac{Q(t)}{2}=0 \,.
\ee
This equation can be rewritten as 
\be
\label{Qa}
\frac{1}{a^3} \frac{d}{dt}\left( \alpha a^3\right) +\frac{Q(t)}{2} 
=0 \,,
\ee
which integrates to 
\be
\alpha(t) =-\frac{1}{2a^3(t) } \int dt \, a^3(t) Q(t) \,.
\ee
Note that in the case $\alpha=0$ the two fluids become perfect and
non-interacting, that is, $Q=0$.

A possible physical interpretation is the following. Fluid~1, 
described by 
$T_{ab}^{(1)}$,  is not  a perfect fluid and its effective energy 
density is not $\rho_1$ but
\be
T_{ab}^{(1)} u^a u^b = \rho_1+2\alpha \,,
\ee
while its effective pressure is still
\be
\frac{1}{3} \, T_{ab}^{(1)} h^{ab}=P_1 \,.
\ee
Fluid~2, instead, has effective energy density and pressure
\be
T_{ab}^{(2)}u^a u^b =\rho_2 -2\alpha \,,
\ee
\be
\frac{1}{3} \, T_{ab}^{(2)} h^{ab}=P_2 \,,
\ee
respectively. In this picture, it would be incorrect to think of 
these two fluids as 
perfect fluids. The terms $\pm \left( q_au_b +q_b u_a \right)$ in 
$T_{ab}^{(i)}$ describe an energy transfer which happens 
simultaneously at all points of space, without transfer of 
three-dimensional momentum, and spoil the perfect fluid nature of 
these fluids. The amount of energy lost by fluid~1 per 
unit time and per unit volume is instantaneously gained by fluid~2, 
and {\em vice-versa}. 
%This situation is different from the one, 
%more familiar in cosmology, in which two fluids are simultaneously 
%present and one comes to dominate the dynamics while the other 
%becomes irrelevant (such as, for example, a radiation fluid with 
%density $\rho_{radiation} \simeq a^{-4}$ and a dust with 
%$\rho_{dust} \simeq a^{-3}$ --- these radiation and dust fluids 
%are 
%decoupled). 
This picture provides the underlying explanation for the splitting
(\ref{fluid1})-(\ref{fluid2}) in the standard approach, where an 
energy transfer occurring 
simultaneously at all points of space is introduced by hand.

Alternatively, one could describe our situation as follows: when $\alpha>0$,
the correction $2\left( \dot{\alpha}+3H\alpha \right)$ to the 
perfect fluid
part of fluid~1 can be visualized as a dust with zero pressure and energy
density $2\alpha$ which supplies energy to fluid~1, while taking it from
fluid~2 through an immediate transfer. From the point of view of fluid~2, one
can think of a perfect fluid from which a dust with negative energy density
$-2\alpha$ removes energy to transfer it to fluid~1. Clearly, this second
dust would violate the weak energy condition, but this is not a significant
problem since a similar case occurs in the standard imperfect fluid, where
a purely spatial heat flux density $q^c$ describes a spacelike, instantaneous
transfer of energy which violates the energy conditions and is clearly 
unphysical, but is still useful as a toy model for a consistent 
relativistic
theory without all its complications. In this sense, the model described by 
eqs.~(\ref{fluid1}) and (\ref{fluid2}) may indeed be acceptable as  
a phenomenological toy model. 

Finally, note that the usual quantity $Q(t)$ introduced in the literature
is related to $\alpha (t)$ through (\ref{Qa}) as 
\be
Q(t)=-\frac{2\left( \alpha a^3 \right)\dot{} }{a^3} \,.
\ee
The physical meaning becomes apparent  if we consider a region of 
three-dimensional
space with unit comoving volume and physical volume $a^3$. Then $-2\alpha
a^3$ is just the energy transferred between the two fluids in this volume, 
$-(2\alpha a^3 )\dot{}$ is the rate  at which this energy transfer 
occurs, and $Q(t) $ is the rate at which this energy is transferred 
per unit volume.

\section{A Lagrangian description}
\label{sec:3}

Having constructed a consistent covariant description of the two-fluid
interaction, the question that arises naturally is whether these equations
can arise from a Lagrangian. The disadvantage is that the two fluids
are not perfect and thus, as it is well known, there is not a robust
Lagrangian formulation for imperfect fluids. Additionally, there is not even
a consensus on how one should proceed in order to approach it. If such a
Lagrangian density is found, it could be possible to give a Lagrangian and
covariant description of two interacting fluids.

Given the high degree of symmetry of FLRW geometry, for inspiration one can
proceed along the lines of a less-known treatment of the classical
dissipative oscillator \cite{oscillator}, in which the oscillating position
$x(t)$ of a point particle is ruled by the usual equation
\be\label{1Doscillator}
\ddot{x}+2\gamma \dot{x} +\omega_0^2 x=0 \,,
\ee
where $\gamma$ and $\omega_0$ are positive constants. 
The change of variable $x(t) = \mbox{e}^{-\gamma t} q(t)$ transforms the
equation of motion (\ref{1Doscillator}) into the new equation
\be
\ddot{q} +\omega^2 q=0 \,,
\ee
where $\omega\equiv \sqrt{\omega_0^2- \gamma^2}$, which is 
dissipationless \cite{oscillator}. The change of variable $x(t) 
\rightarrow q(t) $ is a canonical transformation which makes the 
system Lagrangian, with Lagrangian function
\be
L \left( q, \dot{q} \right)= 
\frac{\dot{q}^2}{2}-\frac{\omega^2 q^2}{2},
\ee
which does not depend explicitly on time. The momentum 
conjugated to 
$q$ is 
\be 
p\equiv \frac{dL}{d\dot{q}} = \mbox{e}^{\gamma t} \left( 
\dot{x}+\gamma \, x \right) 
\ee
and the associated Hamiltonian is
\be
{\cal H}= p\dot{q}-L=\frac{p^2}{2} +\frac{\omega^2q^2}{2} \,.
\ee
Much has been written on this way of removing dissipation from the 
oscillator, the physical interpretation of this procedure and of 
the new Hamiltonian variables $\left( q, p \right)$, and on 
possible quantizations of the dissipative oscillator 
\cite{oscillator}. 

Let us now be inspired by the above treatment, and try to follow the general
idea of removing dissipation from the physical system by changing the
variables which describe the motion. In particular, since we desire to remove
the combination $q_au_b +q_bu_a$ from the stress-energy tensor 
(\ref{bbb}), we should redefine the spacetime variables themselves, that is
redefine the metric. Since $q^c=\alpha u^c$, we can transform the metric
$g_{ab}\rightarrow \bar{g}_{ab}$ according to the Kerr-Schild transformation
\cite{KerrSchild, KSMcH}
\be \label{gbar}
\bar{g}_{ab}=g_{ab}+2 \lambda \alpha u_au_b 
\ee
where $\lambda$ is a constant with the dimensions of an inverse 
density, thus
making the product $\lambda \alpha$ and the metric $\bar{g}_{ab}$
dimensionless (the condition $\lambda \geq 0$ 
guarantees that the metric $\bar{g}_{ab}$ has the same signature 
as $g_{ab}$). The inverse metric straightforwardly reads
\be
\bar{g}^{ab}=g^{ab}+\frac{2\lambda \alpha}{2\lambda \alpha-1} \, 
u^a  u^b,
\ee
and is defined for $ \alpha \neq \frac{1}{2\lambda}$. We restrict ourselves 
to this case: in the pathological situation $\alpha=\frac{1}{2\lambda}$ 
the metric $\bar{g}_{ab}$ degenerates into the 3-dimensional metric 
$h_{ab}\equiv g_{ab}+u_a u_b$ with Euclidean signature, which cannot describe 
the full spacetime metric.

For the spatially flat FLRW spacetime, the line element would 
become 
\be
d\bar{s}^2 =  -\left[1-2\lambda \alpha(t) \right] dt^2 +a^2(t) 
\left( dx^2+dy^2+dz^2 \right) \,.
\ee
It can then be transformed back to the form 
$ d\bar{s}^2 =  - d\bar{t}^2 +a^2(t) d\vec{x}^2 $ by the 
redefinition of the time coordinate 
\be
\bar{t}(t)=\int dt \sqrt{ 1-2\lambda \alpha(t) }\, .
\ee
The  stress-energy tensor becomes
\begin{eqnarray}
T_{ab} &=& \left(P+\rho \right) u_au_b +Pg_{ab} +q_au_b + q_b u_a 
\nonumber\\
&&\\
&=& 
\left( P-  2\lambda \alpha P+ \rho +2\alpha \right) u_a u_b  
+P\bar{g}_{ab} \,.
\end{eqnarray}
This expression formally describes the stress-energy tensor of a 
perfect  
fluid with energy density 
\be
\bar{\rho} = \rho +2\alpha -2\lambda \alpha P 
\ee
and pressure $\bar{P}=P$ in the spacetime metric $\bar{g}_{ab}$. 
Therefore, this stress energy tensor will be covariantly conserved 
according to the covariant derivative $\bar{\nabla}_c$ {\em of the 
metric} $ \bar{g}_{ab} $ \cite{Wald}, namely
\be
\bar{\nabla}^b T_{ab}=0 \,.
\ee
Thus, since in the metric $\bar{g}_{ab}$ the fluid is perfect, and having
in mind the well-known result that the Lagrangian density of a perfect
fluid is just $\sqrt{-g} \, P$ \cite{Seeliger, Schutz, 
Brown}, we can write
down the Lagrangian density associated with the 
stress-energy tensor $T_{ab}$ as 
\be
{\cal L}=\sqrt{-\bar{g}}  \,P \,,
\ee 
where $\bar{g}$ is the determinant of $\bar{g}_{ab}$.
In summary, the ``dissipative'' term $q_au_b +q_b u_a$ has indeed been 
eliminated from the stress-energy tensor and a Lagrangian 
description has been found for this fluid, but at the   
price of introducing a fictitious metric that depends on that 
particular fluid.

The metric $\bar{g}_{ab}$, in which the imperfect fluid becomes perfect, is
not universal: if two different fluids are considered simultaneously, there
will be two different metrics $\bar{g}_{ab}^{(1)}$ and $\bar{g}_{ab}^{(2)}$
and one cannot give a consistent description of the two fluids in the same
``effective spacetime''. In the case of the two fluids~(\ref{fluid1}) and 
(\ref{fluid2}), the metrics $\bar{g}_{ab}^{(1)}$ and 
$\bar{g}_{ab}^{(2)}$ given by eq.~(\ref{gbar}) with $\alpha$ and 
$-\alpha$ respectively, are different.

The situation is similar to that occurring in the classical mechanics of
point particles, in which one can eliminate the (conservative) forces acting
on a particle, by introducing a fictitious space such that the particle
follows geodesics of an effective metric in this fictitious
space\footnote{If there are a finite number of centers of attraction or
repulsion for a particle, its motion under these forces can be reduced to a
geodesic flow, as mentioned in \cite{Kolmogorov} and proved in \cite{Abe}.} 
--- the Jacobi form of the least action principle \cite{Goldstein}. More
generally, one can remove forces acting on a  particle, or self-interaction 
terms in the equation for a field, by introducing  a fictitious 
metric\footnote{The fictitious metric is obtained by means of a 
conformal transformation in \cite{FOP, BiesiadaRugh, Saa}.} in a  
fictitious space \cite{Abe, FOP, BiesiadaRugh, Saa}. However, 
there is a different effective space for each
particle or field considered, and one cannot consider two (or more)
particles or fields simultaneously in this kind of approach, but only 
self-interactions (this statement is true also for test fluids, see 
Appendix~A). Nevertheless, the formal result of this section may 
still be 
useful  for a single fluid.

\section{Scalar field fluids}
\label{sec:4}

In this section we desire to go one step further, and investigate the case
where the fluid is the effective description of a scalar field (see
\cite{Vlagrangianfluids} and references therein). Scalars are the simplest
fundamental physical fields, and since there is no shortage of scalar
fields in high energy theories, a scalar field is often used in the cosmology
of the early and late universe. In principle, a scalar field can be
coupled to a fluid or to another  field. Thus, in this section we briefly
discuss a covariant description of this possible coupling.

\subsection{A fluid and  a scalar field}

We begin by considering two coupled fluids in an FLRW universe, the 
first 
being an ordinary fluid with energy density $\rho_1$ and 
pressure $P_1$, and the second fluid arising from a canonical scalar
field $\phi$ minimally coupled to the curvature (which, when decoupled 
from the dust fluid, is equivalent to an effective perfect fluid). We would
like to offer a theoretical justification of the interaction form
\be \label{Bfluid1}
\dot{\rho}_1+3H \left(P_1+\rho_1 \right) =Q \,, 
\ee 
\be 
\dot{\rho}_{\phi}+3H \left(P_{\phi}+\rho_{\phi} \right) =-Q \,. 
\label{conservationphi} 
\ee
The effective energy density and pressure of a fluid arising from a scalar
field in an FLRW space are given by the well known formulas 
\begin{eqnarray} 
\rho_{\phi} &=& \frac{\dot{\phi}^2}{2}+V(\phi) \,,\label{phidensity}\\ 
&&\nonumber\\ 
P_{\phi} &=& \frac{\dot{\phi}^2}{2}-V(\phi) \,.\label{phipressure} 
\end{eqnarray} 
Finally, adding eqs.~(\ref{Bfluid1}) and (\ref{conservationphi}) one obtains
a 
conservation 
equation for the ``total perfect fluid'' characterized by energy density 
$\rho_{tot}=\rho_1+\rho_{\phi}$ and pressure $P_{tot}=P_1+P_{\phi}$.

In order to provide a theoretically justified form of the fluid-field
interaction term, we are inspired by the large amount of research devoted in
the 1980's literature on inflation reheating. In particular, one 
should find
an interaction term as a phenomenological way to describe the decay of the 
inflaton due to its coupling to other particles, a term that would excite 
the production of this particle in order to end inflation after 
the number of e-folds of expansion needed to solve the horizon and 
flatness problems \cite{KolbTurner}.   Later on, the scenarios for 
ending
inflation took a more definite shape in the various works on reheating and 
preheating. Thus, inspired by the inflaton phenomenological interaction we
consider 
\be \label{Q}
Q= \Gamma \dot{\phi}^2 \,,
\ee
with $\Gamma$ a positive constant. Then, using 
eqs.~(\ref{phidensity}) and (\ref{phipressure}), the equation of 
motion (\ref{conservationphi}) for the 
scalar field becomes 
\be 
\dot{\phi} \left( \ddot{\phi}+3H\dot{\phi} +\Gamma 
\dot{\phi} +\frac{dV}{d\phi} \right)=0 
\ee 
and, unless $\phi$ is a constant $\phi_0$ (in which case the scalar field 
fluid reduces to a pure cosmological constant $\Lambda =V(\phi_0)$ 
and decouples from the first fluid), we have a Klein-Gordon 
equation with a  potential and an extra source of ``friction'' with 
strength described by 
$\Gamma$ and proportional to the ``speed'' $\dot{\phi}$ of the scalar, namely
\be 
\ddot{\phi}+3H\dot{\phi} +\Gamma 
\dot{\phi} +\frac{dV}{d\phi} =0 \,. 
\ee 

Correspondingly, the perfect fluid part of fluid~1 enjoys a source 
$\Gamma \dot{\phi}$ in the right hand side of eq.~(\ref{Bfluid1}),  
\be 
\dot{\rho}_1+3H \left(P_1+\rho_1 \right) 
=\Gamma \dot{\phi} \,. 
\ee 
The quantity $\alpha$ introduced in the 
previous section is 
\be 
\label{alphaff}
\alpha(t)= -\frac{\Gamma}{2a^3} \int dt \, 
a^3 \dot{\phi}^2 \,, 
\ee 
and it involves only the kinetic energy 
$\dot{\phi}^2/2$ of the field $\phi$. The decay of 
the field $\phi$ into the fluid is due to its kinetic energy and 
stops if $\phi$ becomes static. Thus, we can apply the procedure of the
previous section, with the above $\alpha$ quantifying the imperfectness,
obtaining a covariant formulation of the fluid-field interaction.

\subsection{Two scalar field fluids}

Now let the first fluid be also a scalar field $\psi$ with 
self-interaction potential $
U(\psi)$. In this case $\rho_1 = \frac{\dot{\psi}^2}{2}+U(\psi)$ and $
P_1 = \frac{\dot{\psi}^2}{2}-U(\psi) $ and the equation of motion for $\psi$ 
becomes
\be 
\ddot{\psi}+3H\dot{\psi} - \Gamma \,\frac{ \dot{\phi}^2}{\dot{\psi}}  
+\frac{dU}{d\psi} = 0  
\ee 
(we assume that $\dot{\psi}\neq 0$ and $\Gamma >0$). Thus, when 
$|\dot{\psi}|$ is large 
(that is, a ``fast-moving'' $\psi$) and increasing, there is a 
comparatively
small extra 
term $- \Gamma \,\frac{ \dot{\phi}^2}{\dot{\psi}}$ which enhances the motion 
of $\psi$ and could perhaps be interpreted as a sort of ``anti-friction'' 
for this field, a force which depends on the velocities of both $\psi$ and 
$\phi$. However, when $\psi$ is decreasing, this term turns into friction 
opposing the motion of $\psi$. Thus, one can also apply the formulation of
the previous section, with $\alpha$ given by (\ref{alphaff}) quantifying the
imperfectness.

\section{Discussion}
\label{sec:5}

The increasing amount of literature on mutually coupled dark energy and dark
matter, and of a scalar field explicitly coupled to other forms of matter in
cosmology \cite{AmendolaTsujikawabook, miscQ, Bertolami, ThomasValerio},
raises the problem of finding a covariant description of the widely used 
formulation of energy exchange between two fluids. In the present work we
have constructed such a covariant formulation, where the interaction is a
natural implication of the imperfectness of the fluids. 

This imperfectness makes difficult the final step towards a robust
formulation of interacting fluids, namely the construction of a Lagrangian,
whose variation would give rise to the interacting equations, since we need
to face the issue of finding Lagrangian descriptions of dissipative systems,
which is notoriously difficult. We have presented a formal solution to this
problem for a single fluid, entailing the introduction of an effective
metric which depends on this particular fluid. However, its applicability
beyond one fluid is limited, since each fluid sees a different effective
metric.

In summary, we have constructed a covariant description for an 
otherwise {\em ad hoc}, coordinate dependent, formalism widely used 
in cosmology, introducing imperfectness. Whether imperfectness is a 
necessary (apart from sufficient) condition for interaction is 
still an open question, however this seems reasonable from the 
microscopic point of view since in general one cannot easily 
imagine an effective sector to be simultaneously ``perfect'' and 
``interacting''. If this is the case, then it will be very hard, if 
not impossible, to construct a Lagrangian formulation in the usual 
way, for the dark energy-dark matter interaction. And vice-versa, 
if the microscopic nature of dark matter and dark energy is some 
day 
understood, their possible interacting terms in the fundamental 
Lagrangian will probably give rise to a different effective 
interacting behavior than the one used in the current literature.

\begin{acknowledgments} 
The research of VF is supported by Bishop's University and by the Natural
Sciences and Engineering Research Council of Canada. The research of ENS is
implemented within the framework of the Action ``Supporting Postdoctoral
Researchers'' of the Operational Program ``Education and Lifelong Learning''
(Actions Beneficiary: General Secretariat for Research and Technology), and
is co-financed by the European Social Fund (ESF) and the Greek State. 
\end{acknowledgments}

\section*{Appendix~A}

Consider a {\em single test fluid} which is not isolated but 
interacts with another system in FLRW geometry according to
eq.~(\ref{fluid1}). 
The effective equation of state parameter of this fluid is defined 
by $w \equiv P/\rho$ and eq.~(\ref{fluid1}) takes the form
\be\label{testfluid1}
\dot{\rho}+3 \left( w+1 \right)H \rho =Q(t) \,.
\ee
We search for a solution of this equation in the form 
\be
\rho(t)=\frac{C(t)}{a^{3(w+1)}(t)} \,.
\ee
Inserting this ansatz in eq. (\ref{testfluid1}) we acquire 
\be
\dot{C}=Q(t) a^{3(w+1)}(t) \,,
\ee
which is immediately integrated to yield
\be\label{solutionTest}
\rho(t) =\frac{C(t)}{a^{3(w+1)}(t)}= \frac{ 
C_0+\int_0^t dt' Q(t') a^{3(w+1)}(t') }{ a^{3(w+1)}} \,,
\ee
where $C_0$ is an integration constant. 

However, note that if two fluids with equation of state 
parameters $w_1$ and $w_2$ interact according to 
eqs.~(\ref{fluid1}) and (\ref{fluid2}), the 
solution~(\ref{solutionTest})  does not 
apply because then, adding these equations term to term, one would 
obtain
\be
\dot{\rho}_1+ \dot{\rho}_2 
+3\left( w_{tot}+1 \right) H\rho_{tot} =0 
\ee
and the test fluid solutions would be 
\begin{eqnarray}
\rho_1 &=& \frac{ 
C_1+\int_0^t dt' Q(t') a^{3(w_1+1)}(t') }{ a^{3(w_1+1)}} \,,\\
\nonumber\\
\rho_2 &=& \frac{ C_2+\int_0^t dt' Q(t') a^{3(w_2+1)}(t') }{ 
a^{3(w_2+1)}} \,.
\end{eqnarray}
In this case $\rho_{tot}=\rho_1+\rho_2$ has a complicated form which does 
not 
correspond to the ``total fluid'' being a perfect fluid (unless 
$w_1=w_2$, which is the trivial case of  a fluid interacting with 
itself, therefore, of  a single fluid). A total perfect fluid 
should instead have 
$\rho_{tot}= \rho_1+\rho_2$ scaling with one well-defined  
power of $a$ equal to $-3(w_{tot}+1)$.

% Create the reference section using BibTeX:
%\bibliography{simplified}

\end{document}